\documentstyle[12pt]{article}
\begin{document}
\title{\bf{Study of Gamow States in the Rigged Hilbert Space
with Tempered Ultradistributions.}
\thanks{\it{This work was partially supported by Consejo Nacional de
Investigaciones Cient\'{\i}ficas, Comisi\'on de Investigaciones
Cient\'{\i}ficas de la Pcia. de Buenos Aires, Argentina and by
PMT-PICT0079 of ANPCYT, (FONCYT) Argentina.}}}
\author{A. L. De Paoli, M. A. Estevez, M. C. Rocca and H. Vucetich.\\
Departamento de F\'{\i}sica, Fac. de Ciencias Exactas,\\
Universidad Nacional de La Plata.\\
C. C. 67 (1900) La Plata, Argentina.}
\date{January 4, 1999.}
\maketitle
\begin{abstract}
In this work we show that it is possible to extend analitically,
and with the use of tempered ultradistributions, the pseudonorm
defined by T. Berggren for Gamow states. We define this pseudonorm
for all states determined by the zeros of the Jost function 
for any short range potential.

As an example we study the s-states corresponding to the
square well potential.


\end{abstract}

\newpage

\renewcommand{\theequation}{\arabic{section}.\arabic{subsection}.
\arabic{equation}}
\section{Introduction}

Resonant states play a central role
in the quantum description of decaying nuclear states. In the ordinary
formulation of Quantum Mechanics these states appear as complex energy
poles of the $S$ scattering matrix. In turn, these states can be
defined as solutions of the time-independent Schr\"{o}dinger equation
with purely out-going waves at large distances \cite{tp4}.
Several attempts have been performed to handle adequately
resonant states, the main obstacle being the divergent behaviour
of the corresponding wave functions at large distances, which makes
it impossible to normalize them in an infinite volume with the
conventional mathematical tools.
The first succesful attempt to handle resonant wave functions
has been made by Tore Bergreen
\cite{tp5} using a regularization method first suggested by Zel'dovich.
In his work Bergreen has shown that at least for finite range potentials
it is possible to define an ortogonality criteria among bound and
resonant states, and also a pseudonorm can be evaluated
using the general analysis of Newton \cite{tp4}.

A proper inclusion of resonant states within the general framework of
Quantum Mechanics has been done through the Rigged Hilbert Space (RHS) or
Gelfand's Triplet (GT) formulation \cite{tp2}. Resonant states are
described, within the RHS, as generalized complex energy solutions 
of a self-adjoint Hamiltonian.
The structure of the RHS guarantees that
any matrix element involving resonant states is a well defined quantity,
provided the topology in the GT has been properly choosen to handle
the exponential growing of Gamow States at large distances.

The literature concerning the aplication of RHS to resonant states is
extensive \cite{tp9}-\cite{tp12}.
Among these works we shall mention, for instance, those of Bohm
\cite{tp9}, Gadella \cite{tp10} and also reference \cite{tp11},
where resonant states are introduced using a RHS of entire
Hardy-class functions defined in a half complex energy-plane. This allows
to extend analyticaly the concept of a resonant state
as an antilinear complex functional over the intersection of
Schwartz test functions with Hardy class.

A more general theory of resonant states follows if the RHS is built up
on tempered ultradistributions \cite{tp14}. In this case resonant states
arise as continuous linear functionals over rapidly decreasing entire 
analytical test functions. 
This can be obtained by using the Dirac's formula, which
allows a more direct determination of these states.
Another advantage of using tempered ultradistributions is that only
the physical spectrum appears in the definition of complex-energy states
\cite{tp14}.

In the present paper we want to show that
it is possible to define a complex pseudonorm for
resonant states in the sense of Bergreen using tempered ultradistributions. 
With this pseudonorm  we generalize 
the Bergreen's result \cite{tp5}, and hence it can be
considered as the proper analytical extension of a pseudoscalar 
product for resonant states.

We give an introduction of tempered ultradistributions and Gelfand
Triplet in section 2. In section 3 we define resonant states starting
from the Schr\"{o}dinger equation, and then we focus our atention
on the calculus of the pseudonorm of a complex-energy state.
We apply in section 4 the results of the previous section to the case
of a square well potential. We give a resume in section 5.

\section{The Tempered Ultradistributions}
\setcounter{equation}{0}
\subsection{The Triplet $(H,{\cal H},{\Lambda}_{\infty})$}
We define the space $H$ of test functions $\phi(x)$ such that 
$e^{p|x|}|D^q\phi(x)|$ is bounded for any $p$ and $q$ by means of the
set of countably norms (ref.\cite{tp1}):
\begin{equation}
{\|\hat{\phi}\|}_p^{''}=\sup_{0\leq q\leq p,\,x} 
e^{p|x|} \left|D^q \hat{\phi} (x)\right|\;\;\;;\;\;\;p=0,1,2,...
\end{equation}
According to the ref.\cite{tp2} $H$ is a space ${\cal K}\{M_p\}$
with:
\begin{equation}
M_p(x)=e^{(p-1)|x|}\;\;\;;\;\;\; p=1,2,...
\end{equation}
\begin{equation}
{\|\hat{\phi}\|}_p=\sup_{0\leq q \leq p}
M_p(x)\left|D^q \hat{\phi}(x) \right|
\end{equation}
${\cal K}\{e^{(p-1)|x|}\}$ satisfies condition $({\cal N})$ of Guelfand
( ref.\cite{tp3} ). Then if we define:
\begin{equation}
{<\hat{\phi}, \hat{\psi}>}_p = \int\limits_{-\infty}^{\infty}
e^{2(p-1)|x|} \sum\limits_{q=0}^p D^q \overline{\hat{\phi}} (x) D^q 
\hat{\psi} (x)\;dx \;\;\;;\;\;\;p=1,2,...
\end{equation}
\begin{equation}
{\|\hat{\phi}\|}_p^{'}=\sqrt{{<\hat{\phi}, \hat{\phi}>}_p}
\end{equation}
${\cal K}\{e^{(p-1)|x|}\}$ is a countable Hilbert and nuclear space.
\begin{equation}
{\cal K}\{e^{(p-1)|x|}\} = H = \bigcap\limits_{p=1}^{\infty} H_p
\end{equation}
where $H_p$ is the completed of $H$ by the norm (2.1.5).
Let 
\begin{equation}
<\hat{\phi}, \hat{\psi}> = \int\limits_{-\infty}^{\infty}
\overline{\hat{\phi}}(x) \hat{\psi}(x)
\end{equation}
Then, the completed of $H$ by (2.1.7) is ${\cal H}$, the Hilbert space
of square integrable functions. Now
\begin{equation}
<\hat{\phi}, \hat{\psi}> \leq C\,{\|\hat{\phi}\|}_1^{'}
\,{\|\hat{\psi}\|}_1^{'}
\end{equation}
and according to ref.\cite{tp3} the triplet
\begin{equation}
\left(H,{\cal H},{\Lambda}_{\infty}\right)
\end{equation}
is a Rigged Hilbert space or Guelfand's Triplet. Here
${\Lambda}_{\infty}$ is the dual of $H$ and it consist of
distributions of exponential type $T$ ( ref.\cite{tp1} ):
\begin{equation}
T=D^p\left[e^{p|x|} f(x) \right] \;\;\;;\;\;\;p=0,1,2...
\end{equation}
where $f(x)$ is bounded continuous.
\subsection{The Triplet $(h,{\cal H},{\cal U})$} 
\setcounter{equation}{0}
The space $h={\cal F}\{H\}$ ( ${\cal F}$= Fourier transform )
consist of entire analytic rapidly decreasing test functions given
by the countable set of norms :
\begin{equation}
{\|\phi\|}_{pn} = \sup_{|Im(z)|\leq n} {\left(1+|z|\right)}^p
|\phi (z)|
\end{equation}
Then $h$ is a ${\cal Z}\{M_p\}$ space, complete and countable 
normed ( Frechet ) with:
\begin{equation}
M_p(z)= (1+|z|)^p
\end{equation}
If we define:
\begin{equation}
{<\phi (z), \psi (z) >}_p={<\hat{\phi}(x), \hat{\psi}(x)>}_p
\end{equation}
then,  ${\cal Z}\{(1+|z|)^p\}$ is a countable Hilbert and nuclear
space. Let be:
\begin{equation}
\psi (z) = \int\limits_{-\infty}^{\infty} e^{izx} 
\hat{\psi} (x) dx
\end{equation}
\begin{equation}
\phi (z) = \int\limits_{-\infty}^{\infty} e^{izx}
\hat{\phi} (x) dx
\end{equation}
\begin{equation}
{\phi}_1(z)=\frac {1} {2\pi} \int\limits_{-\infty}^{\infty}
e^{-izx} \overline{\hat{\phi}}(x) dx
\end{equation}
Then we define:
\[<\phi (z), \psi (z)>=\int\limits_{-\infty}^{\infty}
{\phi}_1(z) \psi (z) dz = 
\int\limits_{-\infty}^{\infty} \overline{\hat{\phi}}(x)
\hat{\psi}(x) dx \]
\begin{equation}
=<\hat{\phi}(x), \hat{\psi}(x)>
\end{equation}
The completed of $h$ by this last scalar product is the Hilbert
space ${\cal H}$ of square integrable functions
and the dual of  $h$ is the space ${\cal U}$ of tempered 
ultradistributions ( ref.\cite{tp1} ). Then $(h,{\cal H},{\cal U})$ is
a Guelfand's triplet.

The space ${\cal U}$ can be characterized as follow (ref.\cite{tp1}).
Let be ${\cal A}_{\omega}$ the space of all functions $F(z)$ such that:

(i)$F(z)$ is analytic in $\{z\in {\cal C} : |Im(z)|>p\}$.

(ii)$F(z)/z^p$ is bounded continuous in $\{z\in {\cal C} : 
|Im(z)|\geq p\}$ where $p$ depends of $F(z)$. Here $p=0,1,2,...$

Let $\Pi$ be the set of all z-dependent polinomials $P(z)$, $z\in {\cal C}$.
Then ${\cal U}$ is the quotient space:
\begin{equation}
{\cal U}= \frac {{\cal A}_{\omega}} {\Pi}
\end{equation}
Due to these properties any ultradistribution can be represented as a
linear functional where $F(z)\in {\cal U}$ is the indicatrix of this
functional (ref.\cite{tp1}):
\begin{equation}
F(\phi)=<F(z), \phi(z)>=\oint\limits_{\Gamma} F(z) \phi(z) dz
\end{equation}
where the path $\Gamma$ runs parallel to the real axis from
$-\infty$ to $\infty$ for $Im(z)>\rho$, $\rho>p$ 
and back from $\infty$ to $-\infty$ for $Im(z)<-\rho$, $-\rho<-p$
( $\Gamma$ lies outside a horizontal band of width $2p$ that contain
all the singularities of $F(z)$ ).

Formula (2.2.9) will be our fundamental representation for a tempered
ultradistribution. An interesting property, according to ``Dirac formula''
for ultradistributions (ref.\cite{tp8}),
\begin{equation}
F(z)=\frac {1} {2\pi i}\int\limits_{-\infty}^{\infty} dt\;
\frac {f(t)} {t-z} 
\end{equation}
is that the indicatrix $f(t)$ satisfies :
\begin{equation}
\oint\limits_{\Gamma} dz\; F(z) \phi(z) =
\int\limits_{-\infty}^{\infty} dt\; f(t) \phi(t) 
\end{equation}
While $F(z)$ is analytic on $\Gamma$, the density $f(t)$ is in
general singular, so that the r.h.s. of (2.2.11) should be interpreted
in the sense of distribution theory.

The representation (2.2.9) makes evident that the addition of a 
polinomial $P(z)$ to $F(z)$ do not alter the ultradistribution:
\[\oint\limits_{\Gamma}dz\;\{F(z)+P(z)\}\phi(z)=
\oint\limits_{\Gamma}dz\;F(z)\phi(z)+\oint\limits_{\Gamma}dz\;
P(z)\phi(z)\]
But:
\[\oint\limits_{\Gamma}dz\;P(z)\phi(z)=0\]
as $P(z)\phi(z)$ is entire analytic ( and rapidly decreasing ),
\begin{equation}
.{}^..\;\;\;\;\oint\limits_{\Gamma}dz\;\{F(z)+P(z)\}\phi(z)=
\oint\limits_{\Gamma}dz\;F(z)\phi(z)
\end{equation}

In the Rigged Hilbert spaces $(\Phi,{\cal H},{\Phi}^{\ast})$ 
is valid the following very important property:

Every symmetric operator $A$ acting on $\Phi$, that admit a self-adjoint
prolongation operating on ${\cal H}$, has in ${\Phi}^{\ast}$
a complete set of generalized eigenvectors ( or proper distributions )
that correspond to real eigenvalues(ref.\cite{tp3}). 

This property is then valid in $(H,{\cal H},{\Lambda}_{\infty})$ 
and in $(h,{\cal H},{\cal U})$.

\renewcommand{\theequation}{\arabic{section}.\arabic{equation}}
\section{The pseudonorm of eigenstates of short range potentials}
\setcounter{equation}{0}
In this paragraph we describe the main properties of the solutions
of the Schr\"{o}dinger equation for a central short range potential
ref.\cite{tp4}. According to this reference the regular 
( ${\phi}_l(k,r)$ ) and irregular ( $f_l(k,r)$ )
solutions for this equation satisfy, respectively, the following
boundary conditions
\begin{equation}
\lim_{r\rightarrow 0}\;(2l+1)!!\;r^{-l-1}{\phi}_l(k,r)=1
\end{equation}
\begin{equation}
\lim_{r\rightarrow \infty}\;e^{ikr}f_l(k,r)=i^l
\end{equation}
Both solutions are related by
\begin{equation}
{\phi}_l(k,r)=\frac {1} {2} ik^{-l-1}\left[f_l(-k)f_l(k,r)-
(-1)^lf_l(k)f_l(-k,r)\right]
\end{equation}
In (3.3) $f_l(k)$ is the Jost function defined by:
\begin{equation}
f_l(k)=k^l{\cal W}\left[f_l(k,r), {\phi}_l(k,r)\right]
\end{equation}
where ${\cal W}[f,\phi]$ is the Wronskian of the two solutions.
The zeros of the Jost function $f_l(k)$ are the bound 
( $Re(k)=0, Im(k)\leq 0$ ), virtual ( $Re(k)=0, Im(k)>0$ ) and resonant
states ( $Re(k)\neq 0, Im(k)>0$ ) ( refs.\cite{tp4,tp5} ). 
With these definitions we are now in position to calculate the pseudonorm
of the above states. 
According to ref.\cite{tp4} the derivative of the Jost function with 
respect to the variable $k$, ${\dot{f}}_l(k)$, satisfies
\[{\dot{f}}_l(k)=l k^{l-1} {\cal W}\left[f_l(k,r), {\phi}_l(k,r)\right]
+ k^l {\cal W}\left[{\dot{f}}_l(k,r), {\phi}_l(k,r)\right] \]
\begin{equation}
+ k^l {\cal W}\left[f_l(k,r), {\dot{\phi}}_l(k,r)\right] 
\end{equation}
In particular, when $k_0$ is a zero of the Jost function then (3.5) takes
the form:
\begin{equation}
{\dot{f}}_l(k_0)= k_0^l {\cal W}\left[ {\dot{f}}_l(k_0,r), 
{\phi}_l(k_0,r)
\right] + k_0^l {\cal W}\left[ f_l(k_0,r), {\dot{\phi}}_l(k_0,r) 
\right] 
\end{equation}  
Due to eq.(3.3) at $k=k_0$ we have the equality:
\begin{equation}
f_l(k_0,r)=C(k_0){\phi}_l(k_0,r)\;\;\;;\;\;\;
C(k_0)=\frac {-2ik_0^{l+1}} {f_l(-k_0)}
\end{equation}
and following the procedure of ref.\cite{tp4} we get:
\[{\dot{f}}_l(k_0)={k_0}^l \lim_{\beta \rightarrow \infty} \left\{
{\cal W}\left[{\dot{f}}_l(k_0,\beta), {\phi}_l(k_0,\beta)\right] 
-\right.\]
\begin{equation}
\left.2k_0 C(k_0) \int\limits_0^{\beta} {{\phi}_l}^2(k_0,r)\;dr \right\}
\end{equation} 
From (3.8) we deduce immediately:
\[\lim_{\beta \rightarrow \infty} \int\limits_0^{\beta}
{{\phi}_l}^2(k_0,r)\;dr=
- \lim_{\beta \rightarrow \infty} 
\frac {f_l(-k_0)} {4ik_0^{l+2}} {\cal W}\left[{\dot{f}}_l(k_0,\beta),
{\phi}_l(k_0,\beta)\right]\]
\begin{equation}
+\frac {{\dot{f}}_l(k_0) f_l(-k_0)} {4ik_0^{2l+2}}
\end{equation}
Now, we want to show that: i) the integral appearing in (3.9) can be defined
as an ultradistribution in the variable $k_0$ and ii) in the limit 
$\beta \rightarrow \infty$, as an ultradistribution in $k_0$, 
the Wronskian ${\cal W}$ vanishes.
With this purpose and according to ref.\cite{tp4} we note that 
$k^l f_l(k_0,r)$=$h_l(k_0,r)$ is an entire analytic function of the 
variable $k_0$ and therefore $k^{l+1} f_l(k_0,r)$ is also too.
Hence $k^{l+1} {\dot{f}}_l(k_0,r)=g_l(k_0,r)$ is an entire 
analytic function of $k_0$. Moreover it has been shown in ref.\cite{tp4}
that $h_l(0,r)=C{\phi}_l(0,r)$.  
And as a consequence we have $h_l(0,0)=g_l(0,0)=0$ because 
${\phi}_l$ has the property ${\phi}_l(0,0)=0$. 

We can write now (3.8) in terms of $g_l(k,r)$ as:
\[{\dot{f}}_l(k_0)=\lim_{\beta \rightarrow \infty}\left\{ 
\frac {{\cal W}\left[g_l(k_0,\beta), {\phi}_l(k_0,\beta)\right]}
{k_0}\right.\]
\begin{equation}
\left.-2k_0^{l+1} C(k_0) \int\limits_0^{\beta} 
{{\phi}_l}^2(k_0,r)\;dr\right\} 
\end{equation} 
But:
\[\lim_{\beta \rightarrow \infty}\oint\limits_{\Gamma}
\frac {{\cal W}\left[g_l(k_0,\beta), {\phi}_l(k_0,\beta)\right]}
{k_0}\phi(k_0)\;dk_0=\]
\begin{equation}
\lim_{\beta \rightarrow \infty}
{\cal W}\left[g_l(0,\beta), {\phi}_l(0,\beta)\right]
\phi(0)
\end{equation}
where $\phi(k_0)\in h$ is an entire analytic test function and the
path $\Gamma$ runs parallel to the real axis from
$-\infty$ to $\infty$ for $Im(k_0)>\rho$, $\rho>0$ 
and back from $\infty$ to $-\infty$ for $Im(k_0)<-\rho$, $-\rho<0$
( $\Gamma$ lies outside a horizontal band that contains
the singularity in the origin ).
Taking into account that $f_l$ satisfies:
\begin{equation}
\frac {d} {dr} {\cal W}\left[{\dot{f}}_l(k,r), f_l(k,r)\right]=
2kf_l^2(k,r)
\end{equation}
it is easy to show that:
\begin{equation}
\frac {d} {dr} {\cal W}\left[g_l(k,r), h_l(k,r)\right]=
2k^2h_l^2(k,r)
\end{equation}
and then
\begin{equation}
\frac {d} {dr} {\cal W}\left[g_l(0,r), h_l(0,r)\right]= 0
\end{equation}
Eq.(3.14) implies that:
\begin{equation}
{\cal W}\left[g_l(0,r), h_l(0,r)\right]= constant
\end{equation}
and from $h_l(0,0)=g_l(0,0)=0$ we obtain:
\begin{equation}
{\cal W}\left[g_l(0,r), h_l(0,r)\right]=0
\end{equation}
This implies that
\begin{equation}
{\cal W}\left[g_l(0,r), {\phi}_l(0,r)\right]=0
\end{equation}
and then we have:
\begin{equation}
\lim_{\beta \rightarrow \infty}\oint\limits_{\Gamma}
\frac {{\cal W}\left[g_l(k_0,\beta), {\phi}_l(k_0,\beta)\right]}
{k_0}\phi(k_0)\;dk_0=0
\end{equation}
As a consequence of (3.18) it results that:
\begin{equation}
\lim_{\beta \rightarrow \infty}
\frac {{\cal W}\left[g_l(k_0,\beta), {\phi}_l(k_0,\beta)\right]}
{k_0}= P(k_0)
\end{equation}
where $P(k_0)$ is an arbitrary polynomial in the variable $k_0$.

Now we have the freedom to select $P(k_0)\equiv 0$, and in this
case (3.9) takes the form:
\begin{equation}
\lim_{\beta \rightarrow \infty} \int\limits_0^{\beta}
{{\phi}_l}^2(k_0,r)\;dr=
\frac {{\dot{f}}_l(k_0) f_l(-k_0)} {4ik_0^{2l+2}}
\end{equation}
where the limit is taken in the sense of ultradistributions.
By definition the pseudonormalized state is:
\begin{equation}
{\psi}_l(k_0,r)={\left[\frac {4i k_0^{2l+2}} 
{{\dot{f}}_l(k_0)f_l(k_0)}\right]}^{1/2}{\phi}_l(k_0,r)
\end{equation}
and it can be thought as a tempered ultradistribution in the
variable $k_0$.
\section{The square well potential}
\setcounter{equation}{0}
We start from the Schr\"{o}dinger equation for the radial component 
${\cal R}_l(r)$ (ref.\cite{tp6,tp7}):   
\begin{equation}
{\cal R}_l^{''}(r)+\frac {2} {r} {\cal R}_l^{'}(r)+\left[q^2-
\frac {l(l+1)} {r^2}\right]{\cal R}_l(r)=0
\end{equation}
($'$ denotes the derivative $d/dr$) and with 
\begin{equation}
q^2=\frac {2m} {{\hbar}^2}\left[E-{\cal V}(r)\right]=
k^2-\frac {2m} {{\hbar}^2} {\cal V}(r)
\end{equation}
where
\begin{equation}
{\cal V}(r)=\left\{\begin{array}{ll}
                   \;0         & for\;\;r>a \\
                   -{\cal V}_0 & for\;\;r\leq a
                   \end{array}\right.
\end{equation}
and 
\begin{equation}
k^2=\frac {2mE} {{\hbar}^2}
\end{equation}
The regular solution is:
\begin{equation}
{\phi}_l(k,r)=\left\{\begin{array}{ll}
                     q^{-l}r\;j_l(qr) & for\;\;r<a \\
                     r\left[A_l\;j_l(kr)+B_l\;n_l(kr)\right]& for\;\;r>a
                     \end{array}\right.
\end{equation}
where $j_l$ and $n_l$ are,respectively, the spherical Bessel and Newmann 
functions. The constants $A_l$ and $B_l$ in (4.5) are:
\[A_l= ka^2q^{-l}\left[k\;j_l(qa)\;n^{'}_l(ka)-q\;j^{'}_l(qa)\;n_l(ka)
\right]\]
\begin{equation}
B_l=ka^2q^{-l}\left[q\;j_l(ka)\;j^{'}_l(qa)-k\;j^{'}_l(ka)\;j_l(qa)
\right]
\end{equation}
The irregular solution $f_l(k,r)$ is given by:
\begin{equation}
f_l(k,r)=\left\{\begin{array}{ll}
                r\left[C_l\;j_l(qr)+D_l\;n_l(qr)\right] & for\;\;r<a \\
                     -ikr\;h_l^{-}(kr) & for\;\;r>a
                     \end{array}\right.
\end{equation}
where $h_l^{-}=j_l-in_l$ is the spherical Hankel function and the constants
$C_l$ and $D_l$ are given by
\[C_l=-ikqa^2\left[q\;h_l^-(ka)\;n_l^{'}(qa)-k\;h_l^{-'}(ka)\;
n_l(qa)\right]\]
\begin{equation}
D_l=ikqa^2\left[q\;h_l^-(ka)\;j_l^{'}(qa)-k\;h_l^{-'}(ka)\;
j_l(qa)\right]
\end{equation}
Using eqs.(3.4),(4.5) and (4.7) we can evaluate the corresponding Jost
function $f_l(k)$:
\begin{equation}
f_l(k)={\left(\frac {k} {q}\right)}^lika^2\left[k\;j_l(qa)\;h_l^{-'}(ka)
-q\;j_l^{'}(qa)\;h_l^{-}(ka)\right]
\end{equation}
We wish to calculate eq.(3.20) for this example in the case $l=0$. 
With this purpose we need the expressions of 
$f_0(-k_0)$ and ${\dot{f}}_0(k_0)$. For this pourpose
we take into account that $f_0(k_0)=0$.
From (4.9) we obtain for $l=0$ :
\begin{equation}
f_0(k_0)=e^{-ik_0a} \left(ik_0\frac {\sin q_0a} {q_0}+\cos q_0a\right)=0
\end{equation}
and
\begin{equation}
{\dot{f}}_0(k_0)=i\frac {q_0^2-k_0^2} {q_0^3}e^{-ik_0a}
\left(\sin q_0a-q_0a\cos q_0a\right)
\end{equation}
where
\[q_0^2=k_0^2+ \frac {2m} {{\hbar}^2} {\cal V}_0\]
Therefore we deduce from (4.10) and (4.11) that :
\begin{equation}
f_0(-k_0)=-\frac {2ik_0} {q_0} e^{ik_0a} \sin q_0a
\end{equation}
\begin{equation}
{\dot{f}}_0(k_0)=i\frac {q_0^2-k_0^2} {q_0^3}\left(1+ik_0a\right)
e^{-ik_0a}\sin q_0a
\end{equation}
If we replace eqs.(4.12) and (4.13) into eq.(3.20) we obtain finally:
\begin{equation}
\int\limits_0^{\infty}{\phi}_0^2(k_0,r) dr = \frac {1+ik_0a} {2ik_0}\;
\frac {q_0^2-k_0^2} {q_0^4}\;{\sin}^2 q_0a
\end{equation}
It should be noted that when $k_0$ corresponds to a bound 
state the integral (4.14) is real and positive. When $k_0$ corresponds
to a virtual state or a resonant state ($Re k_0\neq 0$,
$Im k_0 > 0$) the integral (4.14) is in general a complex number. It is
not surprising since (4.14) is an analytical extension in the sense of 
ultradistributions of the habitual Lebesgue integral. In fact for a 
bound state ($k_0=-i{\kappa}_0, {\kappa}_0>0$) we have
\begin{equation}
\int\limits_0^{\infty}{\phi}_0^2(k_0,r) dr = \frac {1+{\kappa}_0a} 
{2{\kappa}_0}\;
\frac {q_0^2+{\kappa}_0^2} {q_0^4}\;{\sin}^2 q_0a
\end{equation}
which is the well-known norm of the $l=0$ bound state of the
square well. For the $l=0$ virtual state 
($k_0=i{\kappa}_0, {\kappa}_0>0$) we have:
\begin{equation}
\int\limits_0^{\infty}{\phi}_0^2(k_0,r) dr = \frac {{\kappa}_0a-1} 
{2{\kappa}_0}\;
\frac {q_0^2+{\kappa}_0^2} {q_0^4}\;{\sin}^2 q_0a
\end{equation}
and in this case the integral is real.

\section{Discussion}

We have shown here that tempered ultradistributions allow to perform a
general treatment of complex-energy states, incorporating
in a natural way bound and continuum states as well as resonant and
virtual states together, within a more general framework of Quantum
Mechanics, based on the Rigged Hilbert Space formulation.
In this work we have applied this formulation to the specific evaluation
of the complex pseudonorm, showing that the results come out in a more
transparent way, since they are free from regularization schemes.

As an example of the goodness of the procedure introduced in this
paper we give the evaluation of the pseudonorm of virtual and 
resonant s-states for the square-well potential.


\newpage
\begin{thebibliography}{99}

\bibitem{tp1} M. Hasumi: T$\rm{\hat{o}}$hoku Math. J. {\bf 13},
94 (1961).
\bibitem{tp2} I. M. Gel'fand and G. E. Shilov : Generalized
Functions {\bf Vol. 2}. Academic Press (1968).
\bibitem{tp3} I. M. Gel'fand and N. Ya. Vilenkin : Generalized
Functions {\bf Vol. 4}. Academic Press (1964).
\bibitem{tp8} J. Sebastiao e Silva : Math. Ann. {\bf 136},
38 (1958).
\bibitem{tp4} R. G. Newton : J. of Math. Phys. {\bf 1}, 319 (1960).
\bibitem{tp5} T. Berggren : Nucl. Phys. {\bf A 109}, 265 (1968).
\bibitem{tp6} L. I. Schiff : Quantum Mechanics. McGraw-Hill Kogakusha,
LTD (1968).
\bibitem{tp7} H. M. Nussenzveig : Nucl. Phys. {\bf 11}, 499 (1959).
\bibitem{tp9} A. Bohm : J. of Math. Phys. {\bf 21}, 1040 (1980);
{\bf 22}, 2813 (1981).
\bibitem{tp10} M. Gadella : J. of Math. Phys. {\bf 24}, 1462 (1983);
{\bf 24}, 2142 (1983); {\bf 25}, 2481 (1984).
\bibitem{tp11} A. Bohm, M. Gadella, G. Bruce Mainland :
Am. J. of Phys. {\bf 57}, 1103 (1989).
\bibitem{tp12} A. Bohm and M. Gadella : ``Dirac Kets, Gamow Vectors and 
Gel'fand Triplets''. Lect. Notes in Physics {\bf 348}, Springer, Berlin 
(1989).
\bibitem{tp14} C. G. Bollini, O. Civitarese, A. L. De Paoli,
M. C. Rocca : Phys. Lett. {\bf B 382}, 205 (1996);
J. of Math. Phys. {\bf 37}, 4235 (1996).

\end{thebibliography}
\end{document}